\documentclass[10pt]{iopart}
\usepackage[utf8]{inputenc}
\usepackage{graphicx}
\usepackage[version=4]{mhchem}
\usepackage{siunitx}
\usepackage{longtable,tabularx}
\usepackage{color}
\usepackage{lipsum}  
\usepackage[section]{placeins}
\usepackage{multirow}
\usepackage{cite}
\usepackage[colorlinks, citecolor = red, urlcolor = blue]{hyperref}
\usepackage{makecell} 
\usepackage{amssymb} 
\usepackage{wasysym} 
\usepackage[makeroom]{cancel} 
\usepackage{soul} 
\usepackage{xcolor}
\usepackage{lipsum}  
\usepackage{xpatch} 
\usepackage{upgreek} 
\usepackage{wasysym} 
\usepackage{siunitx} %
\usepackage{gensymb} 

\begin{document}

\title[Experimental characterization of thermionic surface cooling in thermionic discharge]{Experimental characterization of thermionic surface cooling in thermionic discharge}

\author{Junhwi Bak$^{1*}$, Albina Tropina$^1$, James Creel$^2$, and Richard B. Miles$^1$}
\address{$^1$ Department of Aerospace Engineering, Texas A$\&$M University, College Station, Texas $77843$, USA}
\address{$^2$ Bush Combat Development Complex, Texas A$\&$M University, Bryan, Texas $77807$, USA}
\ead{junhwib@tamu.edu}

\vspace{10pt}
\begin{indented}
\item[]{17 September 2022}
\end{indented}

\begin{abstract}
In this work, the thermionic cooling effect during thermionic discharges with parallel plate electrodes at $1$~Torr is investigated. Time-resolved observation of electron emission and surface temperature is realized in addition to the typical steady state characterization. Surface cooling by the electron emission, initiated by plasma ignition, is directly captured at its onset and an estimated cooling capacity of $1.6\pm0.2$~MW/m$^2$ is observed. The present work provides experimental evidence of considerable surface cooling achieved by thermionic cooling. This result indicates that thermionic cooling can be a promising thermal protection method at elevated temperatures, such as those encountered by hypersonic vehicle leading edges in flight. 
\end{abstract}

%
\vspace{2pc}

%

%
\ioptwocol

\section{Introduction}
Electron emission from material surfaces occurs by various mechanisms\cite{Zhang2017100Diodes} such as: thermionic emission\cite{Richardson1901OnPlatinum}, field-emission\cite{Simmons1963GeneralizedFilm,Fowler1928ElectronFields}, photoemission\cite{Fowler1928TheFunction,DuBridge1933TheoryPhotoelectrons}, and secondary electron emission\cite{Lieberman2005PrinciplesProcessing}.

Among the mentioned mechanisms, thermionic electron emission is a phenomenon in which energetic electrons bound in a material escape from the surface when the material is heated. The emitted electron current $j_\mathrm{emi}$ from a material at thermal equilibrium is given by the Richardson-Dushman (RD) equation expressed as\cite{Richardson1921TheBodies, Dushman1923ElectronTemperature},
\begin{equation}
    j_\mathrm{emi} = A_\mathrm{R} T_\mathrm{s}^2 \exp{(-\phi_\mathrm{WF}/k_\mathrm{B} T_\mathrm{s})},\label{eq:rd}
\end{equation}
where $A_\mathrm{R}$ is the Richardson constant, $T_\mathrm{s}$ is the surface temperature, $\phi_\mathrm{WF}$ is the work function, and $k_\mathrm{B}$ is the Boltzmann constant. When an electric field $E_\mathrm{c}$ is present at the cathode surface, $\phi_\mathrm{WF}$ decreases by $\Delta \phi =\sqrt{e^3 E_\mathrm{c}/4\pi \varepsilon_0}$ by the Schottky effect\cite{Schottky1914}.
 
Emitted electrons carry energy away which causes the surface to cool. A heat flux rejected by the emitted electrons from the surface\cite{Richardson1903XIII.Conductors} is given by
\begin{equation}
     {q}_\mathrm{the}=j_\mathrm{em}(\phi_\mathrm{WF}+ 2k_\mathrm{B} T_\mathrm{s}).\label{eq:reject}
\end{equation}
During the 1910s, the cooling effect was first observed by measuring a change of resistance of a heated filament when the thermionic current was cycled on and off\cite{Wehnelt1909UberKorper, Cooke1913XXXVI.II, Cooke1913LXII.Bodies}. The thermionic cooling effect has been known and taken into account in the energy balance of various applications involving thermionic emission, e.g., thermionic power generators\cite{Leblanc1964ThermionicVehicles,Touryan1965AGenerator}, arc discharges\cite{Benilov1995AArcs,Baeva2012Two-temperatureArcs,Ng2015,Yatom2017DetectionIncandescence}, hollow cathodes\cite{Cassady2004Lithium-FedTheory,Goebel2021PlasmaCathodes}, semiconductors\cite{Trushin2018TheoryJunction}, and laser ablation\cite{Zhao2013CoulombFluence}. 

However, only a few experimental studies that focus on the cooling effect of thermionic electron emission have been undertaken. Most existing works reporting thermionic cooling have focused on low-temperature applications\cite{Yangui2019EvaporativeHeterostructures, Hishinuma2003MeasurementsGap}, such as at room temperature, where the thermionic cooling effect was observed in a vacuum thermionic configuration. Another example is carbon nanotubes at temperature $\sim1000$~K. With and without thermionic cooling, steady-state sample surface temperatures were compared, and found a linearity between the surface temperature drop and the emission current\cite{Jin2018ThermionicFilms,Jin2021FunctionalizedApplications}. 

A recent engineering application of thermionic cooling is a thermal protection method for hypersonic vehicles leading edge\cite{Kolychev2012ActiveHeating}. Several theoretical and numerical studies have shown that thermionic cooling at the high-temperature surface can provide considerable cooling capacity that can overtake the radiation cooling at the right conditions\cite{Alkandry2014ConceptualVehicles, Uribarri2015, Sahu2022CesiumFlows}. However, yet no experimental demonstration has been made, and attainable cooling capacity in experiment remains unknown. In these applications, the working pressure is no longer in high vacuum and plasmas are present in surrounding environment.

It is noteworthy that although the cooling effect was demonstrated in vacuum thermionic cooling studies, the effect under applications involving a plasma environment with elevated temperature and pressure has been lacking. In thermionic discharges, various additional aspects such as plasma heating, plasma currents, energy absorption, etc., need to be taken into account in order to understand undergoing physics. One of the reasons for having had a few studies may be partially attributed to those aspects for in most applications involving thermionic emission, the energy rejection by the emitted electron itself has not been the main interest. For example, in hollow cathodes\cite{Cassady2004Lithium-FedTheory, Goebel2021PlasmaCathodes}, often used as neutralizers in electric propulsion, the total amount of electrons that can be extracted from the cathode was typically of interest. Another example is power conversion\cite{Rasor1991, Leblanc1964ThermionicVehicles, Touryan1965AGenerator}, the interest of which was focused on extracting electrical power.

In this work, we aim to characterize the thermionic cooling effect in thermionic discharges, which has not been explored to date and could be a future thermal protection method for applications such as hypersonic or re-entry vehicles. Also, in addition to the typical steady-state characterization of thermionic cooling, for the first time and to the best of our knowledge, time-resolved observations of electron emission and surface temperature are made. This directly captures the moment of surface cooling by the electron emission, initiated by plasma ignition, at its onset. The main focus of this work is on thermionic cooling observation and characterization to demonstrate its capability as a high-temperature surface thermal protection technology. Experimental evidence provided in this work could enhance confidence in high-temperature thermionic cooling applications which mostly have been limited to numerical works.

The structure of the present work is as follows. In Sec.~\ref{sec:thecool}, a summary of thermionic cooling in two categories is provided. Then, Section~\ref{sec:setup} details experimental setups for thermionic discharge and measurement systems.
In Sec.~\ref{sec:disc}, the steady-state surface cooling is first characterized in order to compare the characteristics observed in vacuum thermionic cooling. Then, a time-resolved observation that captures the moment of thermionic cooling is discussed together with a two-dimensional (2D) numerical simulation of heat balance in emitter body.  Lastly, we conclude the work in Sec.~\ref{sec:conc}.

\section{Thermionic cooling in various environments}\label{sec:thecool}

Cooling by the emitted electrons has been referred by various terms determined by its applications; thermionic refrigeration\cite{Mahan1994ThermionicRefrigeration,ODwyer2007ThermionicCurrent}, thermionic cooling\cite{Bescond2018ThermionicHeterostructure,Jin2015ThermionicFilms}, evaporative electron cooling\cite{Yangui2019EvaporativeHeterostructures}, and electron transpiration cooling\cite{Alkandry2014ConceptualVehicles}. All refer to the same phenomenon of cooling by emitted electrons. The applications can be categorized into two main groups depending on the system's operation temperatures; here, we set $1000$~K as a criteria that is where the thermionic emission starts being considerable.

In a common vacuum thermionic emission device, the experimental condition is characterized by ultra high vacuum conditions that is less than $5\times10^{-8}$ Torr. Generally, a high voltage is required in this kind of vacuum emission configuration to overcome the space charge limitation governed by the Child-Langmuir law\cite{Child1911DischargeCaO, Langmuir1913TheVacuum}. As emitted electrons is the only source of the current, evaluating the amount of the emitted current is straightforward. In low-temperature applications, a very low work function is required to have a low enough potential barrier to achieve a considerable emission current for cooling. In semiconductor fields, to reduce the barrier, the Schottky effect is often enhanced by using a high voltage over a short gap, such as in a nm scale across the cathode and anode\cite{Hishinuma2003MeasurementsGap}. Also, vacuum tunnelling cooling\cite{Hishinuma2001RefrigerationDesign,Chua2004ThermionicThermodynamics,Chung2011FieldPbTe} or photo-emission\cite{Schwede2010Photon-enhancedSystems} is utilized to enhance the emission. More details on nanoscale cooling can be found in Ref.~\cite{Ziabari2016NanoscaleReview}.

In high-temperature applications, the electron extraction can be achieved via high voltage, like the low-temperature applications, or by plasma ignition. Examples of thermionic cooling with high-temperature vacuum diode are carbon nanotube films\cite{Jin2015ThermionicFilms, Jin2018ThermionicFilms, Jin2021FunctionalizedApplications}. In certain applications where the cathode to anode gap cannot be short, a required voltage can be difficult to achieve in practice. In such cases, the space charge limitation must be overcome by in alternative ways to achieve thermionic cooling (e.g., plasma ignition, which provides a conductive path through which the emitted electrons can escape from the heated surface).

\begin{figure}
    \centering
    \includegraphics[width=0.95\linewidth]{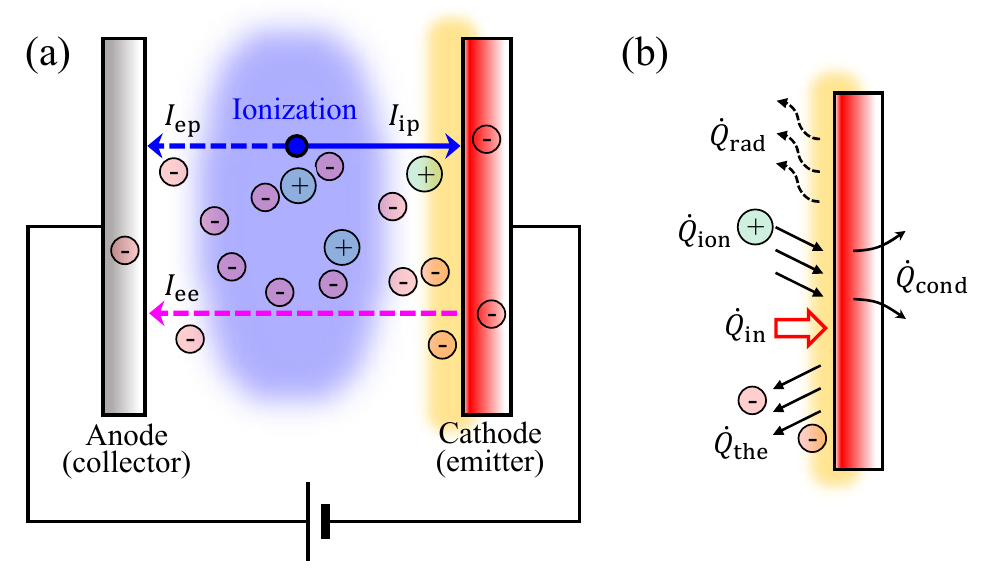}
    \caption{(a) A schematic of thermionic discharge configuration. (b) Energy balance schematic at the emitter front surface in a thermionic discharge.}
    \label{fig:therm}
\end{figure}

Thermionic discharge may be one of the high-temperature applications involving a low-temperature plasma environment for thermionic cooling. A schematic of thermionic discharge configuration is shown in Fig.~\ref{fig:therm}~(a). In thermionic discharges, electron emission occurs under a plasma environment, and plasma provides conductivity throughout the space, allowing the emitted electrons to flow to the anode, avoiding space charge limitation. Further details can be found in Ref.~\cite{Raizer1991GasPhysics}. Known applications of thermionic discharge include power converters\cite{Rasor1991, Lawless1986, Go2017}, neutralizers\cite{Cassady2004Lithium-FedTheory,Goebel2021PlasmaCathodes}, etc.

One recent high-temperature applications of thermionic cooling is thermal protection of high-speed flying vehicles\cite{Kolychev2012ActiveHeating, Kolychev2013ActiveApplicability} such as hypersonic and re-entry vehicles which experience strong aerodynamic heating. For example, hypersonic vehicles can experience very high aerodynamic heating (e.g. a heat influx of $ {q}_\mathrm{in}\approx 3$ MW/m$^2$ for a vehicle with a $10$~mm leading-edge radius at Mach~$8$ and flight at $30$~km altitude\cite{Anderson2019HypersonicEdition}) and require proper a thermal protection to prevent material failure that can lead to the failure of the flight mission. In such atmospheric applications, the distance between electrodes could be far, e.g., from centimeter to even meter scale, compared to the aforementioned low-temperature applications in nm scale. However, surrounding gas particles in the atmosphere can be ionized or dissociated to form plasma, which is expected to mitigate the space charge so that the thermionic cooling can be attained. The thermionic cooling in such applications is often called "electron transpiration cooling (ETC)." In this work, we refer the thermionic cooling as the ETC hereinafter.

It is noteworthy that characterization of thermionic emission under a plasma environment is not trivial due mainly to the plasma itself, e.g., coupling of plasma generation, sheath and electron emission, influence on heat balance, and multiple current components. Sheath potential formation influences the surface electric field of the electrodes, altering the effective work function by the Schottky effect. Surface heat balance also becomes complicated due to additional surface heat influx from plasma, Fig.~\ref{fig:therm}~(b). Note that ion heat influx and electron heat rejection occur simultaneously; thus, the current flow involves multiple components.

\section{Experimental setup}\label{sec:setup}
\subsection{Planar thermionic discharge setup}
\begin{figure}
    \centering
    \includegraphics[width=1\linewidth]{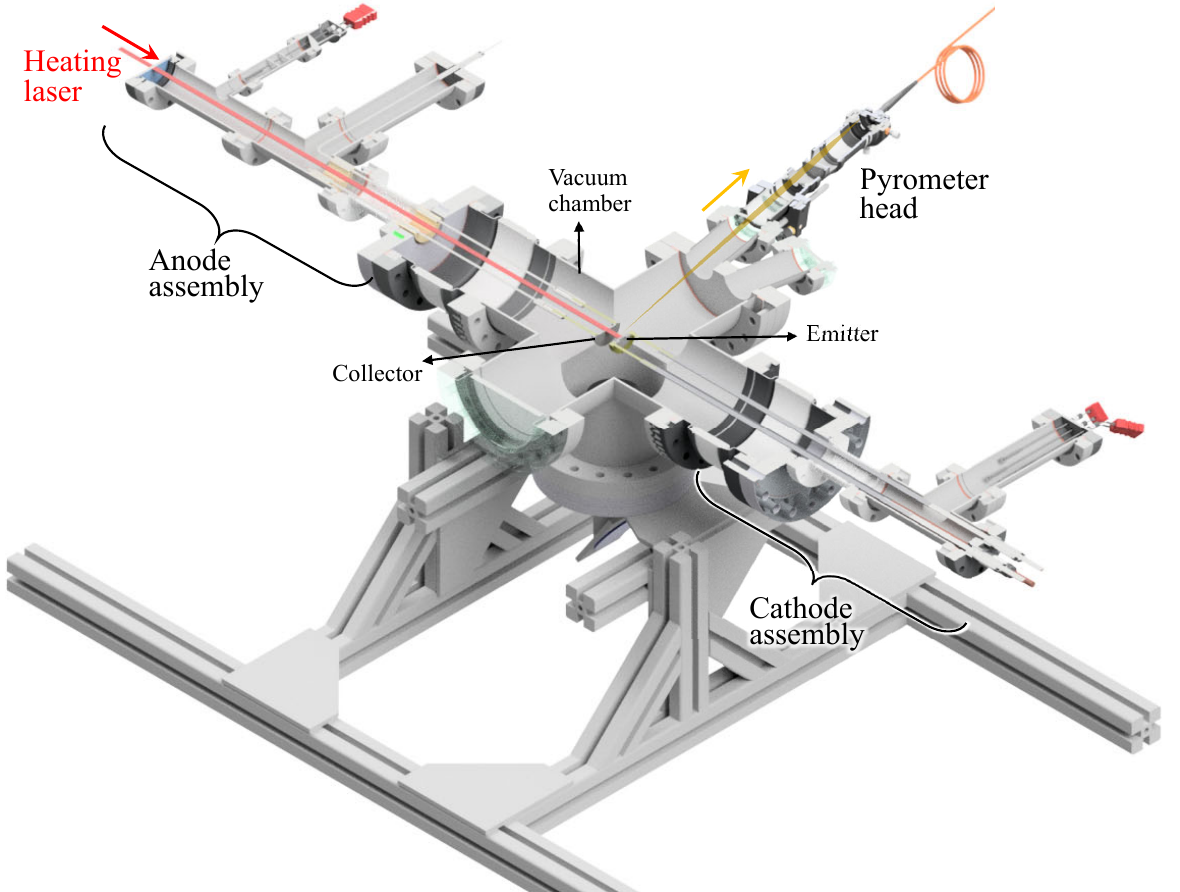}
    \caption{Thermionic cooling test chamber assembly.}
    \label{fig:chamber}
\end{figure}

A plate-to-plate thermionic discharge source was constructed in a vacuum chamber made from a $6$-way $6$-inch diameter ConFlat flange, Fig.~\ref{fig:chamber}. In order to heat the cathode, a $1070$~nm continuous wave (CW) Ytterbium fiber laser (YLR-2000, IPG Photonics) was used as the heating source. The laser can deliver a maximum of 2 kW power, and heat influx into the surface can be rapidly controlled via laser cutoff or modulation at sub millisecond rates. A bias voltage across the electrodes was applied by a DC power supply (KLN $80$-$180$, KEPCO), which accelerates emitted electrons to ignite the plasma. As a background gas, argon was used and maintained at a pressure of $1$~Torr monitored by a capacitance manometer ($10$~Torr range, Type 722A, MKS Baratron). Additionally, xenon and helium were used to observe any gas dependency of thermionic cooling.

\begin{figure}
    \centering
    \includegraphics[width=1\linewidth]{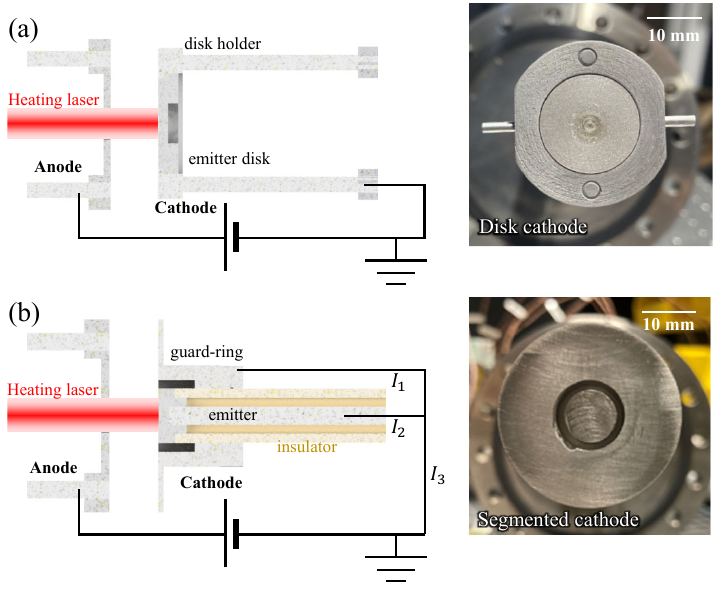}
    \caption{Thermionic discharge electrodes configuration; (a) disk cathode, and (b) segmented cathode.}
    \label{fig:elec}
\end{figure}

Figure~\ref{fig:elec} shows two electrode configurations used for thermionic discharge. The disk cathode, Fig.~\ref{fig:elec}~(a), is the base electrode configuration used for the experiment, and the segmented cathode, Fig.~\ref{fig:elec}~(b), is used to estimate the pure emission current within the total collected current through the circuit. For the disk cathode experiment, lanthanated tungsten, which can have a work function around $2.5$ - $2.7$~eV depending on a lanthanum coverage\cite{Gellert1994,Tanaka2005}, was used. For the segmented cathode, pure tungsten was used, instead, because lanthanum migration in the lanthanated tungsten became problematic, e.g. shorting the segmented parts. Also, we note that the pure tungsten operation did not show notable difference compared to the lanthanated tungsten operation, suggesting that a surface coverage of lanthanum to have a lower effective work function\cite{Gellert1994} is not sufficiently achieved.
A clamp-on current probe (TCP305A, Tektronix) was used to measure the total current, and a current sensor (LES 6-NP, Lam) was used to measure the guard-ring current ($I_2$ in the segmented cathode, Fig.~\ref{fig:elec}~(b)). A bias voltage across the electrodes was measured by a voltage differential probe (DP10013, Micsig). 
 
\subsection{Time-resolved surface temperature measurement}
Careful attention was made to directly monitor the emitter surface temperature, which is needed for time-resolved thermionic cooling characterization. Note that a thermocouple cannot be physically positioned on the surface as it would be affected by the heating laser directly and would also perturb the plasma. If it is positioned behind the front surface of the cathode/anode, the thermal response of the material body cannot quickly transmit a temperature variation on the surface due to its thermal response timescales, i.e., the characteristic time of thermal diffusion is $\sim100$~ms for $2$~mm tungsten\cite{Bak2022StudyModulation}. 

For the time-resolved measurement, the surface temperature was measured by two-color pyrometry\cite{Khan1991NoncontactTechniques, Muller2001DevelopmentEmissivities}. As this is an optical measurement technique, it is non-intrusive and can directly monitor the electron-emitting surface temperature. Pyrometers are often used in fusion research where the plasma facing surface temperature needs to be characterized\cite{Amiel2012SurfacePyrometry, Zhang2017SurfaceTokamaks}. A pyrometer is especially ideal for a time-resolved measurements as fast photodiodes with only a few ns rise time are available. 

The principle of temperature measurement by the pyrometer is based on Plank's radiation law which describes continuous thermal spectral radiation of a physical body. For a body with a spectral emissivity $\varepsilon_\lambda$, the spectral radiance $B$ for wavelength $\lambda$ at temperature $T$ is given by,
\begin{equation}
    B(\lambda,T) = \frac{\varepsilon_\lambda C_1 }{\lambda^5 \left[ \exp(C_2/\lambda T)-1 \right]},
\end{equation}
where $C_1 \equiv 2 h c^2$ and $C_2 \equiv h c / k_\mathrm{B}$ are Planck's radiation constants.

A measured monochromatic signal intensity $I(\lambda,~T)$ of a spectral pyrometer at a wavelength $\lambda$ is then given by $I = K B$, where $K$ is a constant taking into account all parameters related to the signal collection design. The signal ratio $S_{12}$ of the two mono-chromatic signals at two different wavelengths $\lambda_1$ and $\lambda_2$, $S_1$ and $S_2$ respectively, is given as,
\begin{equation}
    S_{12}(T) \equiv \frac{I(\lambda_1,T)}{I(\lambda_2,T)} = \frac{K_1}{K_2}\frac{\varepsilon_{\lambda_1}}{\varepsilon_{\lambda_2}}\frac{\lambda_2^5}{\lambda_1^5} \exp{ \frac{C_2}{T} \left(  \frac{1}{\lambda_2} - \frac{1}{\lambda_1} \right) }, \label{eq:ratio}
\end{equation}
where the Wien approximation, $\exp(C_2/\lambda T) \gg 1$, is used. 
With a calibration, the term $\frac{K_1}{K_2} \frac{\varepsilon_{\lambda_1}}{\varepsilon_{\lambda_2}}\frac{\lambda_2^5}{\lambda_1^5}$ is replaced by the calibration coefficient of the pyrometer $C_\mathrm{pyro}$. Then, the surface temperature $T$ is obtained by the measured intensity ratio $S_{12}$ by the following relation.
\begin{equation}
   T =  \left( A \ln\frac{S_{12}}{C_\mathrm{pyro}}  \right)^{-1}, \label{eq:temp}
\end{equation}
where $A=\frac{1}{C_2} \frac{\lambda_1 \lambda_2}{\lambda_1-\lambda_2}$. 
The pyrometer was calibrated at a steady state condition that covers the operation temperature range by a type-C thermocouple. The thermocouple was attached 2 mm behind the front surface for the disk cathode, and the outer edge of the center emitter for the segmented cathode. It should be noted that, based on a steady-state thermal calculation, approximately $100$-$150$~K systematic error exists due to the offset of the position; however, the random error is suppressed within $\pm2$~K, assuring low-noise temporal measurement of surface temperature. 

\begin{figure}
    \centering
    \includegraphics[width=1\linewidth]{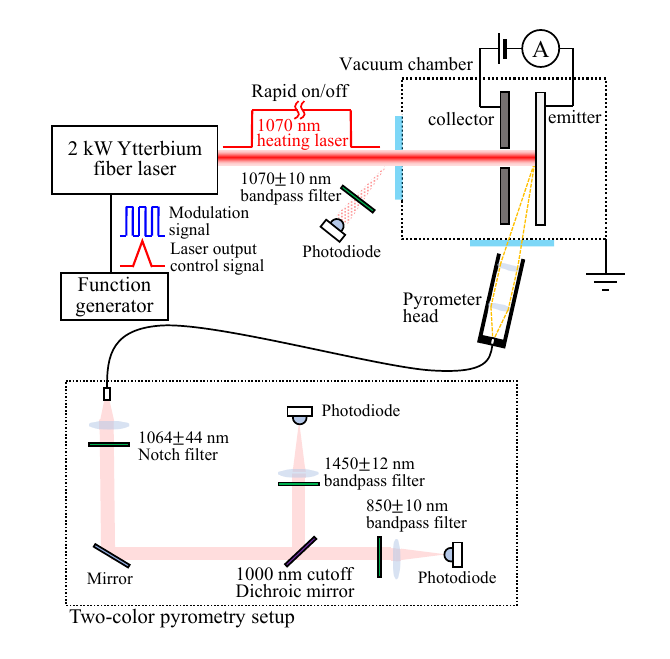}
    \caption{Schematic of a two-color pyrometry setup.}
    \label{fig:pyro}
\end{figure}

In the current pyrometer system, Fig.~\ref{fig:pyro}, thermal radiation of the front surface is collected by a pyrometer head.  The pyrometer head has a $f=250$~mm collimation lens and a $f=100$~mm focusing lens that is coupled to a $2$~m length multimode 400 $\upmu$m core fiber. The output from the fiber is first filtered by a notch filter ($1064\pm44$ nm) which filters out any residual scattering of the $1070$ nm heating laser and split by a dichroic mirror with a $1000$ nm cutoff. The two split components are filtered by narrow bandpass filters, $850\pm10$ nm and $1450\pm12$ nm, and are focused on the fast photodiodes, a Si photodiode (ThorLabs DET100A2, $35$~ns rise time) and an InGaAs photodiode (Thorlabs DET10C2, $10$~ns rise time), respectively. These two wavelengths correspond to the two monochromatic pyrometer wavelengths $\lambda_1$ and $\lambda_2$, assuming the central wavelengths of the filters as representative wavelengths.

\begin{figure}
    \centering
    \includegraphics[width=1\linewidth]{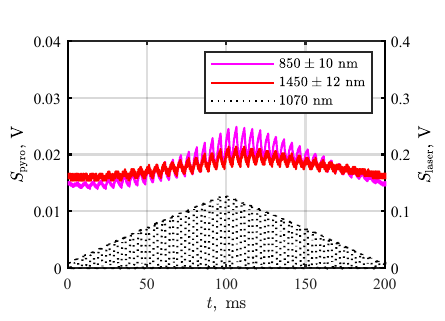}
    \caption{Sample pyrometer data ($850\pm10$~nm and $1450\pm12$~nm) during the surface heating with a $1070$~nm heating laser at modulation frequency $200$~Hz and a ramp frequency $5$~Hz.}
    \label{fig:Pyro_setup}
\end{figure}

Sample raw pyrometry data to confirm fast time-resolved temperature measurements are shown in Fig.~\ref{fig:Pyro_setup}(b). In this test measurement, the heating laser (black dashed) was modulated at the modulation frequency of $200$~Hz with a ramp frequency of $5$~Hz. The heating laser signal is the partial scattering of the heating laser from the inlet optical window that is captured by a photodiode (ThorLabs DET10A2) with a bandpass filter at $1070\pm10$ nm. All photodiode signals were recorded by a high-definition 12-bit oscilloscope (WaveSurfer 4104HD, Teledyne LeCroy). Observed pyrometer responses (magenta for $850$~nm and red for $1450$~nm) confirm a rapid low-noise repeatable temperature measurement of the quickly varying surface temperature.

\section{Result and discussion}\label{sec:disc}
\subsection{Thermionic discharge}
\begin{figure}
    \centering
    \includegraphics[width=.95\linewidth]{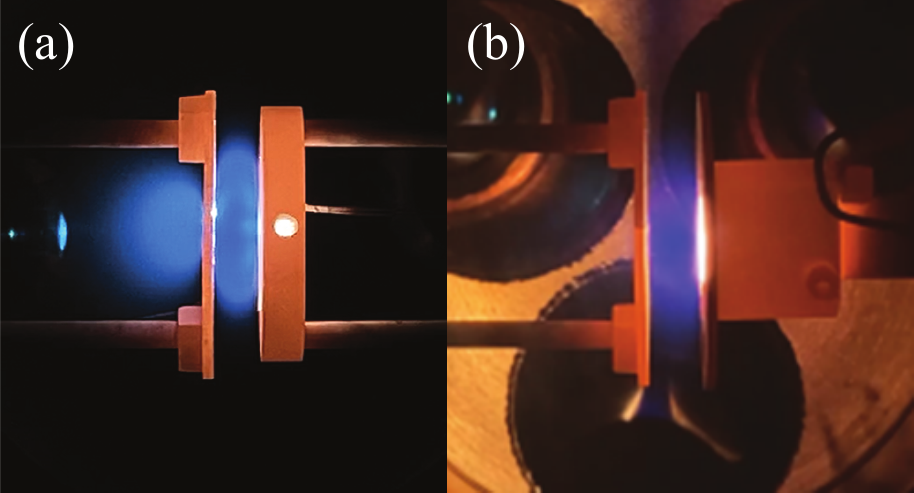}
    \caption{Sample images of thermionic discharge with argon at $1$~Torr with (a) the disk cathode, and (b) the segmented cathode}
    \label{fig:disc}
\end{figure}

Images of thermionic discharge ignited are shown in Fig.~\ref{fig:disc} and the plasma ignition process is described as follows. The front surface of the cathode (the emitter) is heated by the heating laser. At an elevated temperature, electrons having energy that exceed the surface potential barrier are thermionically emitted from the front surface. However, without plasma ignition, they are trapped in the vicinity of the surface due to the space-charge forming a virtual cathode enhancing the potential barrier. Yet, there exists a few electrons at the high energy tail of the energy distribution that could escape the space charge region; however, their energy is not enough to ionize surrounding gas particles. A bias voltage allows such electrons that escape the virtual potential barrier to gain enough energy from the external field. These further heated electrons serve as seed electrons to ionize surrounding gas particles which causes plasma ignition. When the plasma ignites, ions flowing to the emitter release the space charges, allowing the trapped or emitted electrons to flow toward the collector. Several discharge modes in thermionic discharges are well summarized in Ref.~\cite{Malter1951StudiesDischarge}. According to the mode description in the reference above, the present experiment's operation regime with plasma ignition corresponds to the temperature-limited mode\cite{Johnson1956ExternallyArcs}. Note that the electron emission can be turned on or off by controlling the bias voltage, which turns on/off of ETC. Thus, in following sections, "ETC on/off" is achieved by controlling the bias voltage.

\subsection{Steady-state thermionic cooling characterization}
\begin{figure}
    \centering
    \includegraphics[width=0.9\linewidth]{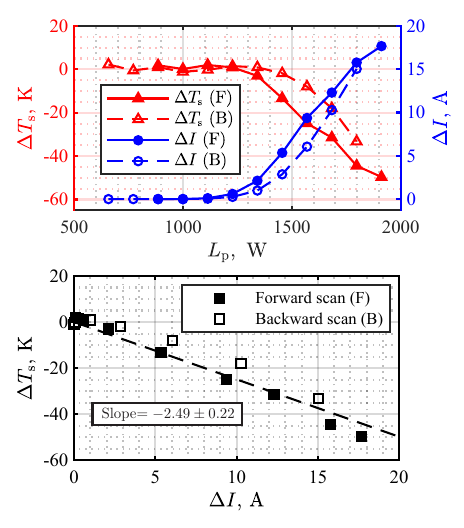}
    \caption{(Top) Surface temperature drop and the total current difference between the cases of ETC on/off as a function of the heating laser power. Data, while laser power is increased or decreased, are noted as 'forward scan (F)' or 'backward scan (B),' respectively. (Bottom) The surface temperature drop as a function of the total current. The linear slope is obtained as $-2.49\pm0.22$.}
    \label{fig:steady}
\end{figure}

Thermionic cooling characterization has often been done by correlating the temperature drop and the emission current, thus, comparing two steady-state conditions with/without thermionic emission\cite{Jin2015ThermionicFilms}. Similar to past work, surface temperature drops $\Delta T_\mathrm{s}$ were examined as a function of the difference in collected current between with/without ETC in Fig.~\ref{fig:steady}. In Fig.~\ref{fig:steady}, the top figure shows the surface temperature drop $\Delta T_\mathrm{s}$ and the total current difference $\Delta I$ between the cases of ETC on/off as a function of the heating laser power. The different laser power yields a higher steady-steady surface temperature that emits more electrons, thus more $\Delta T_\mathrm{s}$, when the ETC is turned on. Data taken while the laser power was increased or decreased are noted as 'forward scan (F)' or 'backward scan (B),' respectively. In the bottom figure, the surface temperature drop is plotted as a function of the total current. The linear slope of $-2.49\pm0.22$ is obtained, which exhibits the linear relation of $\Delta I$ and $\Delta T_\mathrm{s}$. This linear relation agrees with the behavior of the thermionic cooling observed in vacuum emission experiments as well\cite{Jin2015ThermionicFilms,Jin2021FunctionalizedApplications}, which also agrees with the relation $ {Q}_\mathrm{the}=I_\mathrm{emi}(\phi_\mathrm{WF}+ 2k_\mathrm{B} T)$. 

\begin{figure}
    \centering
    \includegraphics[width=0.9\linewidth]{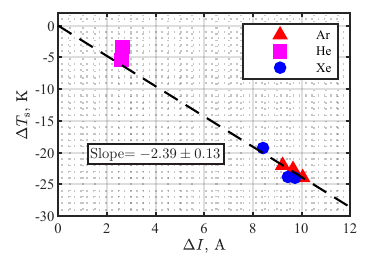}
    \caption{Surface temperature drop as a function of the total current for three different discharge gases; Ar, He, and Xe. The linear slope is obtained as $-2.39\pm0.13$.}
    \label{fig:gas}
\end{figure}

To confirm that the observed behavior in the thermionic discharge setup is not due to any discharge gas-specific phenomena, we tested thermionic cooling under three different background gases (argon, helium, and xenon) at $1$ Torr. Figure~\ref{fig:gas} shows the surface temperature drop between the ETC on-off for Ar, He, and Xe as a function of the collected current. The slope of $-2.39\pm0.13$ agrees well within the error with the previously obtained slope $-2.49\pm0.22$ during the laser power scanning. This agreement confirms that the cooling capacity observed is gas-independent. This is expected as the cooling capacity is determined only by the amount of emitted current and the surface work function.

The small current observed with a He plasma may be attributed to a lower steady-state surface temperature which is expected due to a stronger convection cooling by the helium gas. Note that the thermal conductivity~$k$ of Ar, He, and Xe are $17.7$, $155.7$, and $5.5$ in mW/(m.K)\cite{LemmonThermophysicalSystems}, respectively. Under an assumption of free convection on a vertical surface\cite{Incropera2006FundamentalsTransfer} at $1$~Torr, the convection heat transfer coefficients~$h$ are obtained as $0.8$ for Ar, $2.5$ for He, and $0.5$ for Xe, suggesting the lower surface temperature with He. In the experiment, it was indeed confirmed that the He cases had an approximately $10$~K lower initial steady-state temperature. Additionally, the amount of the extracted current can also be related to different sheath conditions of discharges from gases, thus, different Schottky effects. For instance, the helium which has a higher ionization potential compared to the other gases results in a lower density plasma which may not fully release space-charged limited electrons. Detailed plasma characterization under thermionic discharge will be reserved for a future work.

Further consideration of other possible influences on operation and data collection such as photoemission, plasma optical thickness, and inverse Bremsstrahlung were evaluated and summarized in \ref{appx:other}. Calculations show that all these influences are negligible in our experimental conditions, so do not alter conclusions attained by the observed results.

It should be noted, in the steady-state characterization, although evidence of thermionic cooling is observed, it is difficult to estimate the cooling capacity due to the broad temperature distribution over the bulk emitter body especially since the heat capacity governs the steady-steady temperature. Note that the electron emission is localized at the surface high-temperature region. To fill the gap, time-resolved characterization is made for the localized portion of the emitter on a short timescale.

\subsection{Time-resolved observation of surface thermionic cooling}

\begin{figure}
    \centering
    \includegraphics[width=1\linewidth]{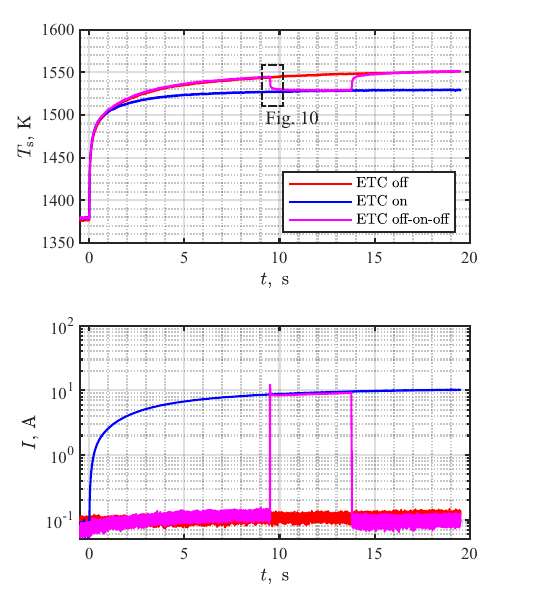}
    \caption{(Top) Time-resolved surface temperature for the ETC off (red), the ETC on (blue), and the ETC off-on-off (magenta). (Bottom) Time-resolved total current for the three cases.} 
    \label{fig:etc_exp}
\end{figure}

In the thermionic discharge, the thermionic cooling can be manually turned on and off by adjusting the bias voltage since the ignition of plasma determines whether emitted electrons can flow across the electrodes. Figure~\ref{fig:etc_exp} shows three experimental cases depending on the on/off of ETC, which were obtained in a similar manner used in Ref.~\cite{Bak2022ExperimentalSystem}. When the cathode is heated and is at steady state conditions, an additional stepwise increase in the heating laser power, $\Delta P_\mathrm{L}\approx640$~W, is applied at $t=0$~s. The top figure shows the temporal evolution of surface temperature, and the bottom figure shows the total collected current.

In the ETC-off case, the bias voltage was kept at $0$~V, which does not ignite the plasma; thus, the electron emission is space-charge limited. However, the bias voltage in the ETC-on case was kept at $15$~V, ensuring plasma ignition. As a result, continuously increasing current history is notable as the temperature rises. Lastly, in the third ETC off-on-off case, the bias voltage starts at $0$~V and is set at $15$~V at $t\approx9.5$~s and set back at $0$~V at $t\approx13.8$~s, showing ETC cycling.

A comparison of the ETC-on and ETC-off cases confirms that continuous electron emissions yield lower surface temperature due to thermionic cooling. The off-on-off case clearly captures the transition between two conditions. Most importantly, the rapid temperature drop is observed at the moment of plasma ignition which corresponds with the current surge. This time-resolved observation is direct experimental evidence of surface cooling by thermionic electron emission.

In this short timescale, the temperature change in a minute element $\delta m$ on the surface from the ignition moment $t'$ can be assumed to be solely the result of the thermionic cooling term $ {Q}_\mathrm{the}$ as other terms are in balance, thus,
\begin{eqnarray}
    \mathrm{at}~t < t'; ~~~~~~~~~~~~0&\approx&\Sigma  {Q}\\
    \mathrm{at}~t \gtrapprox t'; \delta m  c_\mathrm{p} \frac{dT}{dt}&=& {Q}_\mathrm{the} + \Sigma  {Q} \\
    &\approx& {Q}_\mathrm{the} + \Sigma  {Q}_{t<t'} \nonumber \\
    &\approx& {Q}_\mathrm{the}. \nonumber
\end{eqnarray}
where $\Sigma {Q}$ includes the surface heat influx $Q_\mathrm{in}$, the radiation loss $ {Q}_\mathrm{rad}$, and the conduction loss $ {Q}_\mathrm{cond}$.

In order to quantify the cooling rate, the effective mass and heat capacity of the emitting element needs to be determined.  The element mass $\delta m$ of the front surface is given as,
\begin{equation}
    \delta m = A\cdot dx \cdot \rho,
\end{equation}
where $A$ is an emitting surface area, $dx\equiv\sqrt{\eta \cdot dt}$ is the distance that the heat diffuses within the material for $dt$, and $\rho$ is the material density. The thermal diffusivity is given by $\eta\equiv\frac{\kappa}{\rho \cdot c_\mathrm{p}}$ where $\kappa$ is the thermal conductivity and $c_\mathrm{p}$ is the heat capacity. It is assumed that in a millisecond scale, any change in energy balance directly yields the change in the temperature of the minute element. Note that for $dt=6$~ms, the thermal diffusion in tungsten is $dx\approx450~\upmu$m which is approximately $15\%$ of the radius of the laser-irradiated surface. 

The temperature $T_{t+\Delta t}$ of the element after a short time advance $\Delta t$ can be expressed as follows.
\begin{equation}
    \delta m c_\mathrm{p} (T_{t+\Delta t}-T_t)= {Q}_\mathrm{the} \Delta t,
\end{equation}
\begin{equation}
   T_{t+\Delta t}=T_t + \frac{ {q}_\mathrm{the} \cdot \Delta t}{ \rho \cdot c_\mathrm{p} \cdot \sqrt{\eta \cdot \Delta t} },
\end{equation}
where $ {q}_\mathrm{the}\equiv  {Q}_\mathrm{the}/A$. Using the tungsten properties\cite{Lassner1999}, we can estimate $ {q}_\mathrm{the}$ yielding the experimentally resolved $T_\mathrm{s}(t)$.

\begin{figure}
    \centering
    \includegraphics[width=0.95\linewidth]{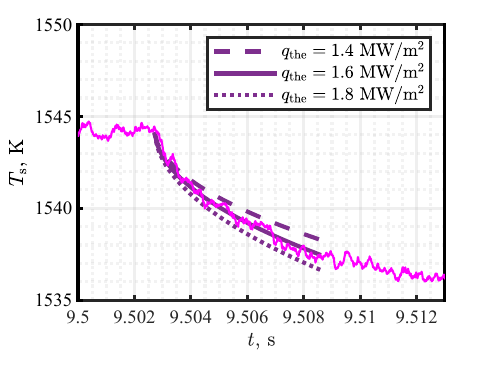}
    \caption{Near the ETC activated moment is zoomed in. Three surface temperature drops depending on the thermionic cooling capacity are overlapped.} 
    \label{fig:etc_zoom}
\end{figure}

Surface temperature history near the ETC activated moment is expanded in Fig.~\ref{fig:etc_zoom}, where three temperature histories depending on the thermionic cooling capacity $ {q}_\mathrm{the}$ are overlapped. Experimentally observed thermionic cooling capacity is estimated to be $ {q}_\mathrm{the}=1.6\pm0.2$~MW/m$^2$.

\subsection{Composition of the total current in thermionic discharge}
It should be recalled that with plasma generation, the total current that is measured in the experiment with the disk-type electrodes is not solely the thermionic emission current. Therefore, in order to investigate the composition of the total current, the segmented cathode is used, Fig.~\ref{fig:elec}~(b). By measuring the two current components, $I_1$ (ion current into the outer segment guard-ring) and $I_3$ (total current), the ion current $I_\mathrm{ion}$ and the emission current $I_\mathrm{emi}$ were obtained. The assumptions that were made are as follows: i) current densities are constant within the center and guard ring areas and ii) no electron emission current is present in the guard ring, but both emission and ion currents are present in the center emitter segment. Then, 
\begin{align}
    I_1 &= I_\mathrm{ion,gr} = j_\mathrm{ion} A_\mathrm{gr}  \label{eq:I1}\\
    I_2 &= I_\mathrm{ion,ctr} + I_\mathrm{emi}=j_\mathrm{ion} A_\mathrm{ctr} + j_\mathrm{emi} A_\mathrm{ctr}\label{eq:I2}\\
    I_3 &= I_1 + I_2 = I_\mathrm{emi} +I_1 A_\mathrm{tot}/A_\mathrm{gr}, \label{eq:I3}
\end{align}
where $A_\mathrm{tot}=A_\mathrm{gr}+A_\mathrm{ctr}$.
From Eq.~(\ref{eq:I1}-\ref{eq:I3}), $I_\mathrm{ion}$ and $I_\mathrm{emi}$ can be expressed as,
\begin{align}
    I_\mathrm{ion} &= I_1 /\alpha \\
    I_\mathrm{emi} &= I_3 -  I_1 /\alpha,
\end{align}
where $\alpha \equiv A_\mathrm{gr}/A_\mathrm{tot}$. From the designed surface area, $\alpha\approx0.9$.

\begin{figure}
    \centering
    \includegraphics[width=0.95\linewidth]{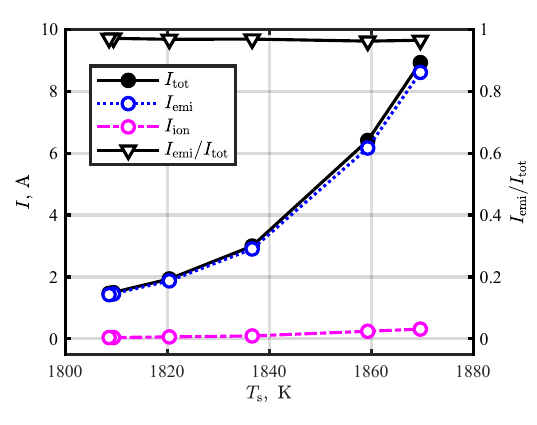}
    \caption{Composition of total current at various surface temperatures. In the present thermionic discharge system, the electron emission current consists of $>96\%$ of the total current.}
    \label{fig:cur}
\end{figure}

Figure~\ref{fig:cur} shows $I_\mathrm{emi}$, $I_\mathrm{ion}$, and $I_\mathrm{tot}$ together with the ratio of $I_\mathrm{emi}/I_\mathrm{tot}$. For the covered total current range, $I_\mathrm{emi}$ is the dominant component of the total current, accounting for $>96\%$ of the $I_\mathrm{tot}$. Thus, it justifies the assumption that the current observed through the disk electrodes is the thermionic current. Note that if the plasma current were a dominant component, the surface temperature would have increased rather than reduced, as the ion flux acts as a heating source. Thus, in this thermionic discharge, the thermionic cooling overwhelms the ion heating for the surface heat balance. This indicates that plasma's role in the surface heat balance is negligible; however, it plays an important role in releasing the space charge, allowing the emitted electrons to escape from the surface. 

\subsection{Two-dimensional numerical simulation of emitter body}

In order to study the ETC effect in more detail, we developed a 2D in-house code to calculate the temperature distribution and heat transfer in the multi-sectional (multi-material) solid body heated by laser irradiation. The solved transient heat conduction equation has a form 
\begin{equation}
    \rho c_\mathrm{p} \frac{\partial T}{\partial t} = \nabla\cdot\left( \kappa \nabla T \right) + q, \label{eq:cond}
\end{equation}
where $\rho$, $c_\mathrm{p}$, and $\kappa$ are the material-dependent density, heat capacity, and thermal conductivity, which are all functions of temperature and space. For the tungsten component, thermal properties were taken from Ref.~\cite{Desai1984,Lassner1999} and, for the molybdenum component material properties were taken from Ref.~\cite{Desai1984,Worthing1926PhysicalTemperature,Rasor1960ThermalTemperatures,Chase1998NIST-JANAFTables}. The heat source term in Eq.~(\ref{eq:cond}) is due to Joule heating $q=j^2/\sigma$, where $\sigma$ is the electrical conductivity of the material and $j$ is the current density. Thus, we considered a problem of electron transpiration cooling of the surface, heated by a laser. We solved the transient heat conduction Eq.~(\ref{eq:cond}) in cylindrical coordinates, using the finite difference method with the second order central scheme for the conductive term and the forward Euler explicit scheme to advance in time. 

\begin{figure}
    \centering
    \includegraphics[width=1\linewidth]{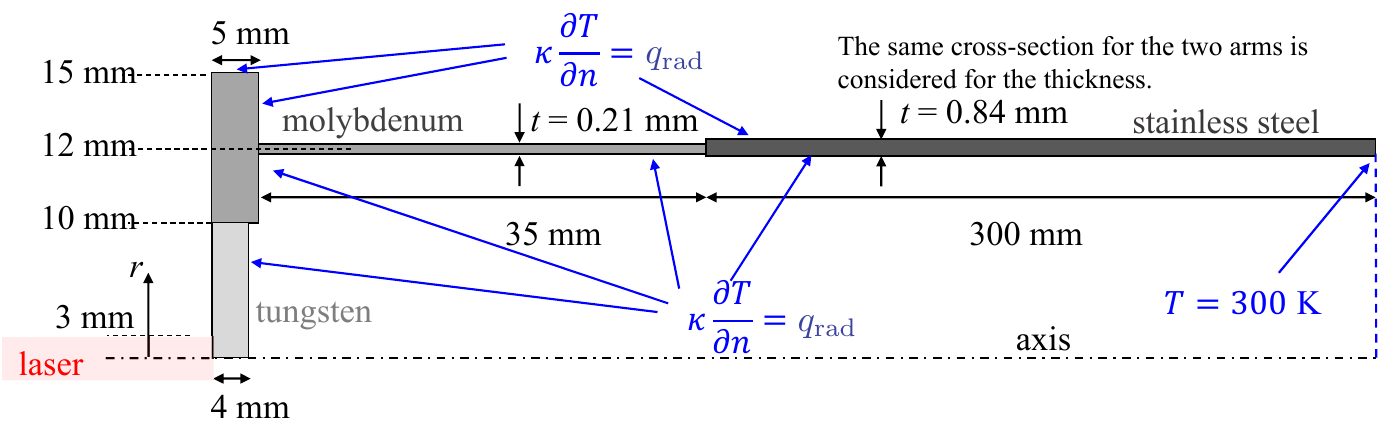}
    \caption{Main geometrical parameters and boundary conditions.}
    \label{fig:sim_con}
\end{figure}

The model geometry with the solid tungsten-molybdenum front surface is shown in Fig.~\ref{fig:sim_con}. Boundary conditions at all outer surfaces are $\kappa \frac{dT}{dz}=q_\mathrm{rad}$, except the right outer surface where we used a Dirichlet boundary condition $T=300$~K and a frontal surface where the thermal flux is given
\begin{equation}
   \kappa \left.\frac{dT}{dz}\right\vert_{z=0} = q_\mathrm{abs}+q_\mathrm{pla}-q_\mathrm{rad}-q_\mathrm{the}. \label{eq:front}
\end{equation}

In Eq.~(\ref{eq:front}) the laser heating term $q_\mathrm{abs}=\beta I_\mathrm{max}\exp(-r^2/(r_\mathrm{spot}^2))$, where $I_\mathrm{max}$ is the maximum laser intensity, $\beta$ is the absorption coefficient and $r_\mathrm{spot}$ is the laser spot radius. Assuming that the maximum spectral density of radiation from the plasma is always limited by the spectral density of the black body radiation and with the temperature increase, the spectrum closely approaches the black body spectrum\cite{Baksht1997RadiationRange}. For the term describing plasma heating near the surface, we use the following relation $q_\mathrm{pla}=\sigma \frac{16 \Lambda T^3 \nabla T}{3}$ \cite{Dresvin1972PhysicsPlasma}. The ETC cooling term is defined as $q_\mathrm{the}=j_\mathrm{emi} (\phi+ 2 k_\mathrm{B} T)/e$, where $\phi$ is a work function of the surface material, $j_\mathrm{emi}$ is the emitted current density and $k_\mathrm{B}$, $e$ are the Boltzmann constant and the elementary charge, respectively.  Radiation losses are accounted for with $q_\mathrm{rad}=\sigma \varepsilon T^4$, where $\varepsilon=\varepsilon(T)$ is the material and temperature dependent emissivity. We performed a grid convergence test, and based on its result, see \ref{sec:grid}, a grid system of $1501\times71$ was chosen.

\begin{figure}
    \centering
    \includegraphics[width=1\linewidth]{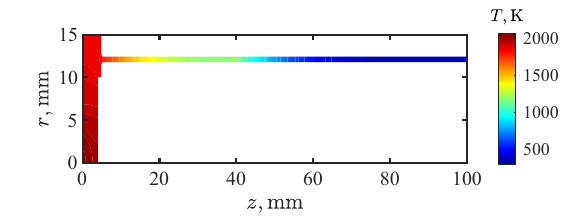}
    \caption{Two-dimensional temperature distribution at a steady-state.}
    \label{fig:sim_ex}
\end{figure}

Simulations were run at conditions corresponding to the experimental conditions. First, a case of steady-state heating of the test article (Fig.~\ref{fig:elec}~(a)) was run with a maximum laser intensity of $I_\mathrm{max}=3.53\times10^7$~W/m$^2$, which corresponds to the experiment laser power of $1000$~W with a laser spot radius of $3$~mm, and an absorption coefficient $\beta$ of 0.12 as an attempt to match the experimental temperature. Sample temperature distribution in the steady state is shown in Fig.~\ref{fig:sim_ex}. It should be noted that the overall temperature in the simulation is higher than that of the experiment. This discrepancy can be attributed to several aspects; the uncertainty in the laser absorption at the surface, the average temperature over the area viewed by the pyrometer, and the temperature calibration location in the experiment. Nonetheless, it is noteworthy that the absolute temperature only affects the balances between heat fluxes terms, i.e., the magnitude of heat fluxes, but does not alter the behavior of ETC dynamics. Here, rather than attempting to match the exact temperature, we focus on evaluating the effectiveness of ETC with varying work functions.

\begin{figure}
    \centering
    \includegraphics[width=1\linewidth]{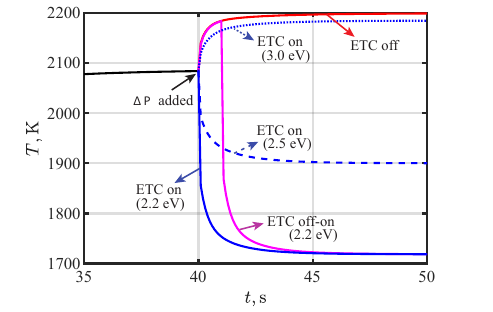}
    \caption{Temporal behaviors of the front surface temperature at $(r,z) = (0,0)$~mm.}
    \label{fig:sim_ETC}
\end{figure}

Using the steady state solution as an initial condition, we ran multiple cases by adding an additional laser power of $\Delta P_\mathrm{L} =670$~W (a maximum laser intensity of $I_\mathrm{max}=5.9\times10^7$~W/m$^2$).  Figure~\ref{fig:sim_ETC} shows the temporal dynamics of the surface temperature at $(r,z) = (0,0)$~mm for three simulated cases: i) ETC-off (red curve); ii) ETC-off, then ETC-on at $41$~s (the magenta curve); iii) ETC-on (blue curves; solid, dashed, and dotted lines correspond to the work function set to $2.2$~eV, $2.5$~eV, and $3.0$~eV, respectively). A temperature decrease is clearly seen when ETC is active, demonstrating surface cooling due to the electron transpiration mechanism. The simulation results reproduce several main experimental observations; the monotonic temperature increase without ETC after the additional energy dump (red solid), the transition behavior between ETC-on and ETC-off (magenta solid), and the moderate temperature rise with ETC-on (blue dotted, $3.0$~eV). Parametric tests with work function show that the surface cooling by ETC can even overwhelm the additionally dumped energy, e.g., $2.2$~eV and $2.5$~eV cases, reaching a lower temperature than the starting temperature. The observed maximum surface temperature drop is about $\sim470$~K with $2.2$~eV. It is noteworthy that such behaviors could only be captured by the controlled ETC activation. 

\begin{figure}
    \centering
    \includegraphics[width=1\linewidth]{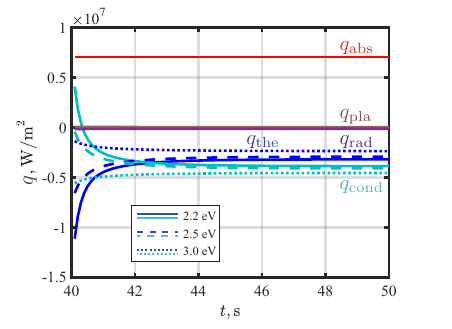}
    \caption{Temporal behaviors of the heat fluxes at $(r,z) = (0,0)$~mm.}
    \label{fig:sim_fluxes}
\end{figure}

Figure~\ref{fig:sim_fluxes} shows the temporal dynamics of the heat fluxes, $q_\mathrm{abs}$, $q_\mathrm{pla}$, $q_\mathrm{rad}$, $q_\mathrm{the}$, and $q_\mathrm{cond}$, at $(r,z) = (0,0)$~mm. Three lines (solid, dashed, and dotted) of $q_\mathrm{the}$ and $q_\mathrm{cond}$ represents the three ETC-on cases with varied material work function, $2.2$~eV, $2.5$~eV, and $3.0$~eV, respectively (as the same manner in Fig.~\ref{fig:sim_ETC}). It is found that $q_\mathrm{the}$ and $q_\mathrm{cond}$ are the two major cooling mechanisms that effectively cancel out the heat influx and that $q_\mathrm{pla}$ and $q_\mathrm{rad}$ are negligible at the calculation conditions. It is interesting to note that, for the case with the work function $2.2$~eV, the initial $q_\mathrm{the}$ cooling flux is so large that $q_\mathrm{cond}$ acts as heating; the conductive heat flows toward the front surface from the surrounding. This is because the ETC is a rapid mechanism that becomes active almost instantly, strongly affecting the heat flux dynamics. When the steady state is nearly reached, the variation in $q_\mathrm{the}$ between three cases stays within $q_\mathrm{the}=2.8\pm0.4$~MW/m$^2$. 

\begin{figure}
    \centering
    \includegraphics[width=1\linewidth]{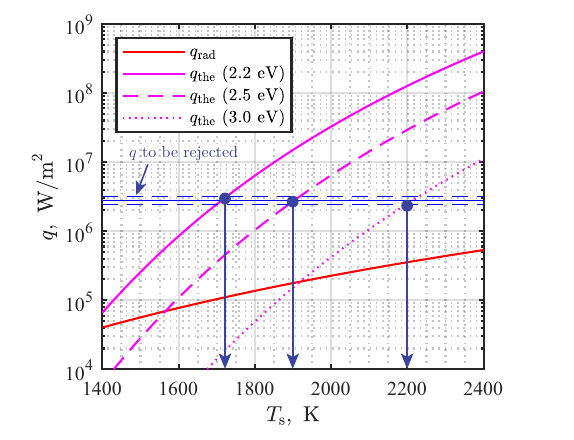}
    \caption{Heat fluxes that can be rejected by $q_\mathrm{rad}$ (red) and $q_\mathrm{the}$ (magenta) as function of the surface temperature. The blue lines show the heat to be rejected. The arrows from the crossing points with $q_\mathrm{the}$ curves show the achievable steady-state surface temperature.}
    \label{fig:qT}
\end{figure}

This heat flux magnitude corresponds to the convective heat flux into a hypersonic vehicle with a $10$~mm leading-edge radius at Mach~$8$ and with a flight altitude of $29$~km. Thus, the steady-state surface temperature results shown can be thought of as the temperature that can be reached when the active ETC is rejecting $q\approx 2.8$~MW/m$^2$. This becomes clear in Fig.~\ref{fig:qT} where heat fluxes of $q_\mathrm{rad}$ (red) and $q_\mathrm{the}$ (magenta) are shown as function of the surface temperature. Note the dominant $q_\mathrm{the}$ over $q_\mathrm{rad}$ at certain temperatures. The blue lines show the heat to be rejected ($q\approx 2.8$~MW/m$^2$). The arrows from the crossing points with $q_\mathrm{the}$ curves show the achievable surface temperatures. These temperaturest match with the converging temperatures seen in Fig.~\ref{fig:sim_ETC}.

\begin{figure}
    \centering
    \includegraphics[width=1\linewidth]{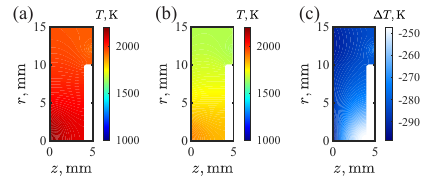}
    \caption{Steady state temperature distributions in the test article. (a) ETC-off case, (b) ETC-on case with $\phi_\mathrm{WF}=2.5$~eV, and (c) the difference between the ETC-off and ETC-on.}
    \label{fig:sim_steady}
\end{figure}

Sample steady-state temperature distributions in the tungsten-molybdenum test article are shown in Fig.~\ref{fig:sim_steady}, where (a) the ETC-off case, (b) the ETC-on case with $\phi_\mathrm{WF}=2.5$~eV, and (c) the temperature difference between the ETC-off and ETC-on. Overall, a $\Delta T \approx -280$~K temperature drop is observed. Comparatively, $\Delta T \approx -470$~K for the $\phi_\mathrm{WF}=2.2$~eV case and $\Delta T \approx -30$~K for the $\phi_\mathrm{WF}=3.0$~eV case. 

\begin{figure}
    \centering
    \includegraphics[width=1\linewidth]{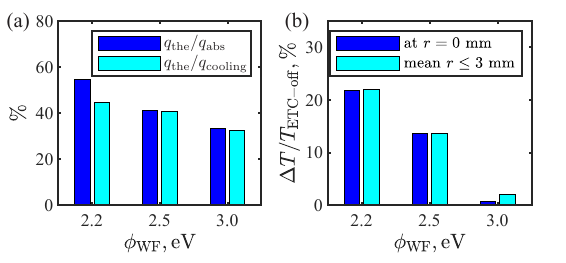}
    \caption{Decrease of the temperature (in K) of the front surface due to the ETC.}
    \label{fig:effi}
\end{figure}

Several steady-state thermal parameters for three work function cases are compared in Fig.~\ref{fig:effi}. Figure~\ref{fig:effi}(a) presents $q_\mathrm{the}/q_\mathrm{abs}$ and $q_\mathrm{the}/q_\mathrm{cooling}$ where $q_\mathrm{cooling}$ is the sum of all cooling fluxes. The direct comparison of the heat fluxes in Fig.~\ref{fig:effi}(a) shows that the ETC can reject considerable portions of the heat influx and still occupy more than $30\%$ of the total cooling flux, suggesting that ETC could be quite effective even with $\phi_\mathrm{WF}=3.0$~eV or higher. However, we need to evaluate how much surface temperature can be actually cooled down by ETC because the whole heat balance process is a non-linear problem. Figure~\ref{fig:effi}(b) shows the surface temperature drop between the ETC-off and ETC-on, $\Delta T$, normalized by the temperature with the ETC-off, $T_\mathrm{ETC-off}$, for the center $r=0$~mm and for the mean over the area within $r\le3$~mm (where the laser heat influx is incident). The results indicate that the actual achievable surface temperature cooling can be insignificant for $\phi_\mathrm{WF}>3.0$~eV, only a drop of a few percent, even though the cooling flux is considerable. Thus, achieving $\phi_\mathrm{WF}<3.0$~eV for a material to be used for an ETC application is paramount to achieving a substantial reduction in temperature.

\section{Conclusion}\label{sec:conc}
The present work successfully established a laboratory thermionic cooling test setup with a laser as a heating source and provided the first experimental evidence of surface cooling under thermionic discharge. Surface cooling by thermionic electron emission in a plate-to-plate thermionic discharge at $1$~Torr was characterized. 

The moment of thermionic cooling activation and corresponding surface cooling was directly captured for the first time indicated by a rapid temperature drop at the transient surge of current when plasma is ignited. The cooling capacity was estimated to be $ {q}_\mathrm{the}=1.6\pm0.2$~MW/m$^2$. It was shown that over 96\% of the total current was found to be the emitted electrons, confirming that the discharge is under the temperature-limited mode and that surface cooling is achieved by the thermionic electron emission. The plasma ion effects from plasma ignition was found to have a negligible effect on the surface heat balance. The main role of plasma ignition is to negate the space charge limitation thereby allowing the emitted electrons to freely escape the surface. 

The observed result indicates that thermionic cooling can be a promising thermal protection method at elevated temperatures, such as hypersonic vehicle leading edge cooling, of which research has been mostly limited to numerical approaches. The present work provides experimental evidence that considerable surface cooling can be achieved by thermionic cooling. It should be noted that the reported cooling capacity is not a limitation of the cooling method. A low work function surface such as a cesium-covered surface\cite{Sahu2022CesiumFlows} should be able to further improve the cooling capability of the method.

The 2D simulation results reproduced experimentally observed ETC dynamics well. The simulation revealed details of heat flux balance at the front surface, showing the dominant cooling by ETC and conduction in the present setup. The parametric tests with varying work functions showed that, although the surface temperature drop in ETC-on with a work function of $3$~eV was $\sim30$~K, a drop of $\sim 300$~K or $\sim 500$~K can be achieved when the work function $2.5$~eV and $2.2$~eV, respectively. From the ratio of the surface temperature drop to that of the non-ETC case, it has been shown that obtaining $\phi_\mathrm{WF}<3.0$~eV for materials for the ETC application is highly important for the effective surface cooling by ETC.

\ack
We gratefully acknowledge support from Lockheed Martin Corporation.

\appendix

\section{Consideration on other influences}\label{appx:other}

\hspace{7mm}\textit{photoemission} - To clarify whether the laser source for surface heating can cause a significant photoemission current from the surface,  we evaluated the photoemission current using the photoemission theory of Fowler and DuBridge\cite{Fowler1931TheTemperatures, DuBridge1932AEmission, Bechtel1977}, as follows
\begin{equation}
    j_n = a_n \left( \frac{e}{h \nu}\right)^n A I^n (1-R)^n T^2 F\left( \frac{nh\nu-\phi}{k_\mathrm{B}T}\right)\label{eq:photo}
\end{equation}
where $R$ is the surface reflectivity, $F$ is the Fowler function,  $h\nu$ is the energy of the incident photon (for the laser wavelength $\lambda_\mathrm{L}$~=~$1.07$~$\upmu$m), $n$ is the number of photons absorbed, $A$ is the  Richardson constant. Equation~(\ref{eq:photo}) at $n$~=~$0$ transforms to the Richardson-Dushman expression for the thermionic emission current. The photoemission current and thermionic current are compared at the laser intensity  $I_L$~=~$5.3\times10^7$~W/m$^2$, which is the laser intensity, evaluated based on our laser characteristics (a laser power $P_\mathrm{L}$~=~$1.5$ kW and a surface area of $0.28$~cm$^2$ from the surface diameter $\diameter$~=~$6$~mm). Parameters for calculating the thermionic current have been chosen as $A$~=~$120\times10^4$~A/cm$^2$/K$^2$, $\phi$~=~$3$~eV.

\begin{figure}
    \centering
    \includegraphics[width=0.9\linewidth]{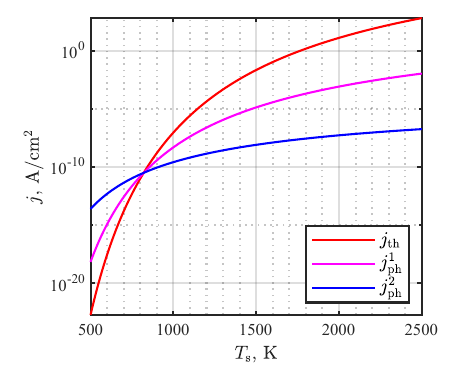}
    \caption{Thermoionic ($n=0$) and photoemission currents as functions of the surface temperature, where $n$ is a number of photons absorbed.}
    \label{fig:photo}
\end{figure}

Figure~\ref{fig:photo} shows the dependence of currents on the surface temperature. It is seen that when the surface temperature is above 900 K, the thermionic current starts to overtake the photoemission current. It means that in the conditions of this experiment, all measured current is current which is due to the thermionic emission. This is indirectly confirmed by the results of the study\cite{Zhou2022TheoryCoating}, where authors used a more advanced quantum approach to find the photoemission current and also took into account laser heating of electrons inside the metal lattice. It is shown in Ref.~\cite{Zhou2022TheoryCoating}, that for the range of the laser intensities we are using in the experiment, the photoemission current is negligibly small.

\textit{Plasma optical thickness} - The influence of ignited plasma on the heating laser is evaluated in terms of the optical thickness. With the laser wavelength $\lambda_\mathrm{L}$~=~$1.07$~$\upmu$m, the laser frequency $\omega_\mathrm{L}\sim10^{15}$~rad/s and the cutoff density $n_\mathrm{cutoff}\sim10^{21}$~cm$^{-3}$ are obtained. For the plasma density $n_\mathrm{e}\sim10^{13}$~cm$^{-3}$, the plasma frequency is $\omega_\mathrm{pe}\sim10^{11}$~rad/s. As $n_\mathrm{cutoff}\gg n_\mathrm{e}$ and the refractive index $n=\sqrt{1-\omega^2_\mathrm{pe}/\omega^2_\mathrm{L}}\approx1$, the plasma is optically thin with respect to the heating laser. 

\begin{figure}
    \centering
    \includegraphics[width=0.9\linewidth]{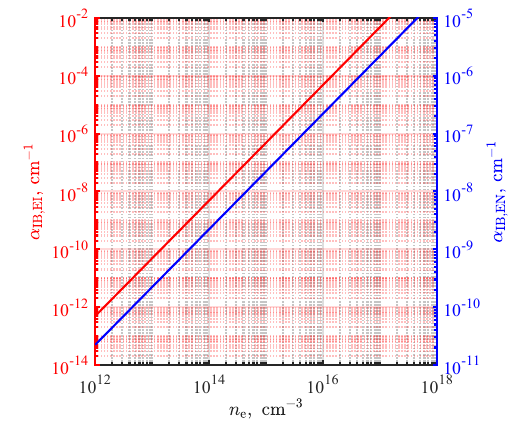}
    \caption{The electron-ion IB absorption coefficient $\alpha_\mathrm{IB,EI}$ and the electron-neutral IB absorption coefficient $\alpha_\mathrm{IB,EN}$ as a function of the plasma density $n_\mathrm{e}$.}
    \label{fig:IB}
\end{figure}

\textit{Inverse Bremsstrahlung} - In the present work, where a laser is used as a heat source and plasma exists, the heating laser power being delivered to the surface can be affected by inverse Bremsstrahlung (IB). The laser energy absorption by the IB can occur by the interaction of electrons with either ions or neutral particles. The electron-ion IB absorption coefficient $\alpha_\mathrm{IB,EI}$ and the electron-neutral IB absorption coefficient $\alpha_\mathrm{IB,EN}$ in cm$^{-1}$ are expressed as\cite{Kemp1980Laser-heatedReport}

\begin{equation}
     \alpha_\mathrm{IB,EI} =  \num{1.37e-27} G \lambda^3 T^{-\frac{1}{2}} n_\mathrm{e} n_\mathrm{i}( e^{0.014388/\lambda T} -1 ), \nonumber
\end{equation}
\begin{equation}
     \alpha_\mathrm{IB,EN} = \num{9.60e-5} T^2 A(T) \lambda^3 n_\mathrm{e} n_\mathrm{n} (1- e^{-0.014388/\lambda T}), \nonumber
\end{equation}
respectively, where a Gaunt factor $G$ is given as,
\begin{equation}
    G=1.04+\num{3.74e-5}T-\num{3.28e-10}T^2 \nonumber
\end{equation}
and a factor $A(T)$ is obtained as $T \cdot A(T)$ as a smoother function of $T$ as explained in Ref.~\cite{Kemp1980Laser-heatedReport}. With plasma parameters of $n_\mathrm{e}$~=~$10^{12}$ - $10^{18}$~cm$^{-3}$, $T_\mathrm{e}$~=~$1$~eV and gas parameters of $p$~=~$1$~torr and $T_\mathrm{g}$~=~$300$~K, $\alpha_\mathrm{IB,EI}$ and $\alpha_\mathrm{IB,EN}$ are obtained in Fig.~\ref{fig:IB}. For $n_\mathrm{e}\sim10^{13}$~cm$^{-3}$ in the present thermionic discharge, $\alpha_\mathrm{IB,EI}\sim10^{-11}$~cm$^{-1}$ and $\alpha_\mathrm{IB,EN}\sim10^{-12}$~cm$^{-1}$, indicating the IB has a negligible influence on the heating laser absorption in the present work. 

\section{Grid convergence check for the 2D heat balance simulation}\label{sec:grid}
\begin{figure}
    \centering
    \includegraphics[width=0.95\linewidth]{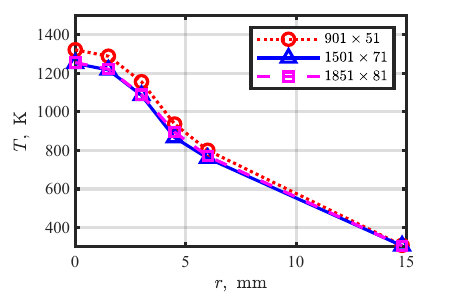}
    \caption{Results of the grid convergence test.}
    \label{fig:grid}
\end{figure}

To verify a code, we performed the grid convergence test. We consider a steady state for the geometry of the test article shown in Fig.~\ref{fig:sim_con}, assuming that the laser spot radius is $3$~mm, delivered power is $1000$~W, and the absorption coefficient $\beta=0.35$. Note that these values are arbitrary for the grid convergence check. Figure~\ref{fig:grid} shows the temperature of the front surface for different grids (the number of grid points in $z$ $\times$ the number of grid points in $r$-direction) : $901\times51$; $1501\times71$; $1851\times81$. We choose the grid $1501\times71$ as a base case because, the further increase of the number of grid points does not significantly change the results as seen from Fig.~\ref{fig:grid}.
\section*{References} 

\bibliography{references.bib}
\bibliographystyle{iopart-num}

\end{document}